\documentclass[a4paper,11pt]{article}

\usepackage{graphicx}
\usepackage[english]{babel} 
\usepackage[utf8]{inputenc}
\usepackage[T1]{fontenc}
\usepackage{dsfont}
\usepackage{lmodern} 
\usepackage{stmaryrd}
\usepackage{amsmath}
\usepackage{amsthm}
\usepackage{amsfonts}
\usepackage{amssymb}
\usepackage{url}
\usepackage[ruled,vlined]{algorithm2e}
\usepackage{color}
\usepackage{multirow}
\usepackage{slashbox}
\usepackage{float}
\usepackage{authblk}

\newcommand{\myhat}{\hat}
\newcommand{\mytilde}{\tilde}

\newcommand{\paren}[1]{\left( \left. #1 \right. \right)} 
\newcommand{\croch}[1]{\left[ \left. #1 \right. \right]} 
\newcommand{\set}[1]{\left\{ \left. #1 \right. \right\}}

\newtheorem{Corollary}{Corollary}

\newtheorem{Lemma}{Lemma}

\providecommand{\keywords}[1]{\textbf{\textit{Index terms---}} #1}

\usepackage[round]{natbib}

\begin{document}

\title{An efficient algorithm for T-estimation}
\author[1,3]{Nelo Magalh\~aes\thanks{nelo.moltermagalhaes@gmail.com}}
\author[2,3]{Yves Rozenholc\thanks{yves.rozenholc@parisdescartes.fr}}
\affil[1]{Équipe Probabilités et Statistiques, Universit\'{e}  Paris-Sud 11}
\affil[2]{MAP5 - UMR CNRS 8145, Universit\'{e}  Paris Descartes}
\affil[3]{INRIA team Select}
\date{September 2014}

\maketitle

\begin{abstract}
We introduce an efficient and exact algorithm, together with a faster but approximate version, which implements with a sub-quadratic complexity the hold-out derived from T-estimation.
We study empirically the performance of this hold-out in the context of density estimation considering well-known competitors (hold-out derived from least-squares or Kullback-Leibler divergence, model selection procedures, etc.) and classical problems including histogram or bandwidth selection. Our algorithms are integrated  in a companion R-package called {\it Density.T.HoldOut} available on the CRAN: {\url{http://cran.r-project.org/web/packages/Density.T.HoldOut/index.html}}. 
\end{abstract}



\section{Introduction} 
Suppose we have at hand a sample of independent and identically distributed (i.i.d.) random variables from some unknown density $s$ with respect to some dominating measure $\mu$ and that we want to estimate $s$ from the sample.

Many papers have been published about the solution of this estimation problem with as little prior information on $s$ as possible. 
A widely used strategy consists in starting from a family of preliminary estimators (for instance kernel or histogram estimators) with some varying smoothing parameter (the bandwidth or the partition) and selecting one candidate using the sample. Nevertheless, since the 30's \citep{Larson1931} it has been known that building estimators and evaluating their quality with the same data yields an overoptimistic result. Many solutions exist to overcome this problem. One natural procedure - called \textit{hold-out} - consists in splitting the sample into two subsamples, building a family of estimators using the first subsample (which we shall call the training sample) and making the selection using the second subsample (which we shall call the validation sample).

Concerning the selection part, \citet[Section 9]{Birge2006b} proposed a procedure 
 - called \textit{T-hold-out} hereafter - based on robust tests between the preliminary estimators. The procedure can be derived from Birgé's construction of T-estimators\footnote{``T'' refers to test.} oriented to model selection. The definition of these estimators is introduced in the same paper but relies on old ideas arising from \citet{LeCam1973,Birge1983,Birge1984a,Birge1984b}. Indeed, conditionally to the training sample, all the estimators are deterministic so that the models are reduced to points and the problem amounts to select one point from the validation sample. \\

The purpose of this paper is to provide an efficient algorithm that implements the T-hold-out, made available in our R-package called \textit{Density.T.HoldOut}. Our motivations are twofold. First, when we  started this research in the summer of 2012 there was no practical application of T-estimation\footnote{recently, Sart has applied robust tests in the special cases of dyadic partition selection \citep{Sart2012} and parameter selection \citep{Sart2013}} and we were very surprised to observe that most authors  - including Birgé himself -  considered this procedure only as a theoretical tool, because of its supposed ``too high computational complexity'' as pointed out in \citet{Birge2006b}, \citet[p.45]{Birge2007} and \citet[p.241]{Baraud-Birge2009}. Second, we thought it would be of interest to compare empirically T-estimation  with classical resampling and penalization procedures since they are motivated by risk estimation, whereas T-estimators are based on robust tests and thus enjoy  some robustness properties.  For this purpose we considered several finite collections of preliminary estimators. These included histogram or kernel collections - leading to some well-known estimation problems: number of bin selection, partition selection, bandwidth selection, but also more complex collections mixing histograms and kernel estimators potentially completed  with some parametric ones. 
{The scripts, developed for this paper using our R-package, are available on the RunMyCode website (\url{http://www.runmycode.org}) to increase transparency and reproducibility}.\\

Hold-out is not specific to the density framework. Indeed, in all cases where we have at hand two independent random samples $\textbf{X}_t$ and $\textbf{X}_v$, one can build a collection of estimators using the training sample $\textbf{X}_t$ and proceed  to the selection with the validation sample $\textbf{X}_v$.  In density estimation, hold-out has been investigated theoretically for projection estimators \citep[Section 8.1]{Arlot-Lerasle2014} and kernel density estimates \citep{Devroye-Lugosi2001} among other examples. Searching for the best linear (or convex) combination of the preliminary estimators in the validation step leads to the linear (or convex) aggregation problem (see \cite{Rigollet-Tsybakov2007}). 
Moreover, theoretical properties of the hold-out have also been studied in classification \citep{Bartlett-Boucheron-Lugosi2002,Blanchard-Massart2006} and in regression -by \cite{Lugosi-Nobel1999,Juditsky-Nemirovski2000,Nemirovski2000,Wegkamp2003}, among others. 

\subsection{Framework}
Let us consider a sample $\textbf{X}=\set{X_{1},\ldots,X_{n}}$ of i.i.d. random variables $X_i$ with values in the measured space $(\mathcal{X},\mathcal{W},\mu)$. We suppose that the distribution of $X_i$ admits a density $s$ with respect to $\mu$ and aim to estimate $s$. We turn the set $\mathcal{S}$ of all probability densities with respect to $\mu$ into a metric space using the Hellinger distance  $h(t,u)$ where 
\begin{equation*} 
h^{2}(t,u) =\frac{1}{2}\int\paren{\sqrt{t(x)} - \sqrt{u(x)}}^{2}d\mu(x)\enspace.
\end{equation*}

Although Birgé's procedure relies on this distance, we shall also consider $L_q$-distances - derived from $L_q$-norms denoted $\|.\|_q$ -  for $q=1,2$.\\

The quality of an approximation $t\in\mathcal{S}$ of the function $s$ is measured by $\ell(t,s)$, where $\ell$ is a loss function (typically some power of a distance). The risk of an estimator $\mytilde{s}=\mytilde{s}(\textbf{X})$ of the function $s$ is defined through this loss function by $R_s(\mytilde{s},\ell):=\mathbb{E}_s[\ell(\mytilde{s},s)]$,
where $\mathbb{E}_s$ denotes the expectation when $s$ obtains. 
The \textit{Hellinger risk}  $R_s(\mytilde{s},h^2)$ comes from the loss $\ell=h^2$. The loss can also be defined as $\ell(t,s)=\mathbb{E}_s[\gamma(t,X)-\gamma(s,X)]$, where 
$\gamma: \mathcal{S} \times \mathcal{X} \mapsto  [0,\infty)$ is  a \textit{contrast function} for which $s$ appears as a minimizer of  $\mathbb{E}_s[\gamma(t,X)]$ when $t\in\mathcal{S} $ \citep[Definition 1]{Birge-Massart1993}. In this context, the $L_2$-loss (resp.\ the Kullback-Leibler loss) is  defined via the contrast function $\gamma(t,x)=\|t\|_2^2-2 t(x)$ (resp.\ $\gamma(t,x)=-\log(t(x))$) for any $t \in \mathcal{S}$, $x\in \mathcal{X}$. 

\subsection{About the Hold-Out}\label{sec: HO}
Formally, the \textit{hold-out} (HO)  is a  two-steps estimation procedure which relies on a split of $\textbf{X}$ into two non-empty complementary subsamples, $\textbf{X}_t$ and $\textbf{X}_v$. 

\begin{itemize}
\item {\bf Step one}: Using the {\it training} sample $\textbf{X}_t$, we build a finite set $S=\set{\myhat{s}_m[\textbf{X}_t], m\in \mathcal{M}}$  of preliminary estimators.
\item {\bf Step two}:  The {\it validation} sample $\textbf{X}_v$ is dedicated to the selection of one point $\myhat{m}$ in $\cal M$. 
\end{itemize} 
The final estimator is either $\myhat{s}_{\myhat{m}}[\textbf{X}_t]$ or $\myhat{s}_{\myhat{m}}[\textbf{X}]$ depending on the authors. The goal is generally to select $\myhat{m} \in \cal M$ such that 
\begin{equation*} 
R_s(\myhat{s}_{\myhat{m}}[\textbf{X}_t],\ell) \sim \inf_{m \in \mathcal{M}} R_s(\myhat{s}_m[\textbf{X}_t],\ell) \quad \text{or} \quad R_s(\myhat{s}_{\myhat{m}}[\textbf{X}],\ell) \sim \inf_{m \in \mathcal{M}} R_s(\myhat{s}_m[\textbf{X}],\ell)\enspace, 
\end{equation*}
where $\ell$ is  the relevant loss function and the symbol $\sim$ means that quantities on both sides are of the same order. \\

Usually, after performing {\it Step one}, one defines some random criterion $\text{crit}(m)$ for each $m$ and selects the $\myhat m\in \mathcal{M}$ that minimizes $\text{crit}(m)$. In the \textit{classical} hold-out, when the loss $\ell$ is defined through a contrast function, this criterion is an estimation of the risk, made using the empirical contrast based on the validation sample:
\begin{equation*}
\text{crit}_{{\rm HO}}(m,\textbf{X}_t,\textbf{X}_v)=\frac{1}{|\textbf{X}_v|} \sum_{X_i\in \textbf{X}_v} \gamma(\myhat{s}_m[\textbf{X}_t],X_i)\enspace,
\end{equation*}
where $|A|$ denotes the  cardinality of the set $A$. In this context one naturally selects the estimator with the smallest estimated risk, 
\begin{equation*}
\myhat m \in \arg \min_{ m \in \mathcal{M}}  \text{crit}_{{\rm HO}}(m,\textbf{X}_t,\textbf{X}_v)\enspace.\end{equation*}

We shall denote in what follows  $\myhat m_{LS}$ and $\myhat m_{KL}$ for the estimators selected by the classical procedure using the contrast functions $\gamma(t,x)=\|t\|_2^2-2 t(x)$ and $\gamma(t,x)=-\log(t(x))$ respectively. We call least-squares hold-out (LSHO) and Kullback-Leibler hold-out (KLHO)  the corresponding HO procedures.
Few theoretical results exist concerning this classical HO in the density framework. Nevertheless, considering projection estimators together with the least-squares contrast, \citet{Arlot-Lerasle2014} have shown that the LSHO criterion can be written as a penalization criterion with some resampling-based penalty.  They also proved an oracle inequality and provided variances computations for this criterion (see Theorem 3 and Section S.2. in the supplementary material in \cite{Arlot-Lerasle2014}). 

\subsection{Overview of the paper}
In practice the selection problem of {\it Step two} amounts to select one estimator in a given collection of $|\mathcal{M}|$ initial candidates. While  the classical HO relies on the optimization of an empirical contrast function and thus requires at most $|\mathcal{M}|$ computations, T-estimation involves pairwise comparisons based on robust tests leading to a quadratic number $O(|\mathcal{M}|^2)$ of tests. 

The first goal of this paper is to provide an algorithm in the general framework of T-estimation which allows an efficient and exact implementation of T-estimation in the HO context. This algorithm breaks this quadratic bound. The second goal is to compare the risk performance of this T-hold-out for two different tests, three losses, a large set of densities and several sample sizes. We shall make a comparison against two types of procedures: those which select one point in a given family using the validation sample and those which estimate the density from the full sample. 

Moreover, we provide a faster, albeit approximate, version of this exact algorithm. We shall study both algorithms from a computational complexity point-of-view as well as the risk performance of the resulting estimators. \\

\noindent
The paper is organized as follows. In Section 2 we revisit the definition of the T-hold-out in a general framework. We introduce in Section 3 our exact and efficient algorithm which implements exact T-estimation and one approximate version derived from it. Section 4 presents the simulation protocol of our empirical study together with a short description of the main function of the companion R-package {\it Density.T.HoldOut}. Section 5 is dedicated to the study of the quality of the two possible T-hold-out in terms of risk. We also provide comparisons with other hold-out procedures, direct estimation procedures --penalized estimators or Lepski's method-- and some bandwidth estimators obtained using asymptotic derivation of the risk. Section 6 is devoted to the empirical study of the complexity of the exact algorithm. Section 7 provides a comparison of exact and approximate algorithms both in terms of risk and complexity.

\section{T-Hold-Out}\label{RHO}

Let us recall the T-hold-out procedure in a general framework where robust tests exist. We have at hand two independent samples, $\textbf{X}_t$ and $\textbf{X}_v$, and want to estimate some target $s$ belonging to the metric space $(\mathcal{S},d)$. Suppose that a family $S=\set{\myhat{s}_m[\textbf{X}_t], m\in \mathcal{M}}$ of estimators of $s$ has been built from $\textbf{X}_t$, and we want to proceed to the selection step with $\textbf{X}_v$. For $m_1, m_2\in \mathcal{M}$, we write $d(m_1,m_2)$ instead of $d\paren{\myhat{s}_{m_1}[\textbf{X}_t],\myhat{s}_{m_2}[\textbf{X}_t]}$. Let us assume that $\psi_{m_1,m_2}$ is a statistical test that decides between $m_1$ and $m_2$ which, conditionally to the knowledge of $S$, is based only on $\textbf{X}_v$. The T-hold-out (THO) criterion is given by
\begin{equation*}
\text{crit}_{{\rm THO}}(m,\textbf{X}_t,\textbf{X}_v):= \sup_{j \in \mathcal{R}_{m}} d(j,m) \enspace,
\end{equation*}
with $\mathcal{R}_{m}$ the set of estimators preferred to $m$, namely $\set{ j\in \mathcal{M}, ~ j\neq m ~|~ \psi_{m,j}=j }$. One finally chooses 
\begin{equation*} \myhat{m} \in\arg \min_{ m \in \mathcal{M}} ~\text{crit}_{{\rm THO}}(m,\textbf{X}_t,\textbf{X}_v)\enspace. \end{equation*}
Considering two densities $\myhat{s}_i[\textbf{X}_t]$ and $\myhat{s}_j[\textbf{X}_t]$, the test is defined by
\begin{equation}\label{eq:psi}
\psi_{i,j}=\left\{
\begin{array}{lll}
        i \quad \text{if} \quad  T_{i,j}\leq 0\\
        \\
        j \quad \text{otherwise}.
\end{array}
\right.
\end{equation}
In the density framework, the test statistic $T_{i,j}$ can be one of the following:
\begin{itemize}
\item setting $\omega=\arccos(1-h^2(\myhat{s}_i[\textbf{X}_t],\myhat{s}_j[\textbf{X}_t]))$,  \citet[Section 4]{Birge2013a} introduced
\begin{equation} \label{test_Birge}
T_{i,j}=\sum_{X_k\in \textbf{X}_v} \log\paren{\frac{\sin(\theta \omega) \sqrt{\myhat{s}_i[\textbf{X}_t]}+ \sin(\omega(1-\theta)) \sqrt{\myhat{s}_j[\textbf{X}_t]}}{\sin(\theta \omega) \sqrt{\myhat{s}_j[\textbf{X}_t]}+ \sin(\omega(1-\theta)) \sqrt{\myhat{s}_i[\textbf{X}_t]}} (X_{k})} \enspace,
\end{equation}
\item setting $r_{i,j}[\textbf{X}_t]=\paren{\myhat{s}_i[\textbf{X}_t]+\myhat{s}_j[\textbf{X}_t]}/2$, \citet[Section 2]{Baraud2011} considered 
\begin{equation}\label{test_Baraud}
T_{i,j}= h^2\paren{\myhat{s}_i[\textbf{X}_t],r_{i,j}[\textbf{X}_t]}-h^2\paren{\myhat{s}_j[\textbf{X}_t],r_{i,j}[\textbf{X}_t]}+\frac{1}{|\textbf{X}_v|} \sum_{X_k\in \textbf{X}_v} \frac{\sqrt{\myhat{s}_j[\textbf{X}_t]}-\sqrt{\myhat{s}_i[\textbf{X}_t]}}{\sqrt{r_{i,j}[\textbf{X}_t]}}(X_k) \enspace.
\end{equation}
\end{itemize}

 To the best of our knowledge it is the first HO  based on the Hellinger distance. There are several theoretical differences with classical HO methods. The criterion  $\text{crit}_{{\rm THO}}(m,\textbf{X}_t,\textbf{X}_v)$ 
does not estimate the risk but appears instead as a \textit{plausibility index}. Its value is computed through robust tests between estimators, while the classical HO criterion is computed independently for each estimator and thus does not take the geometrical structure of $S$ into account. \\

Theoretical results about the THO procedure can be found in \citet[Corollary 9]{Birge2006b} for the Hellinger risk, and in \citet[Corollary 1]{Birge2013b} for the $L_2$-risk. The key assumption in the construction is the existence of some test having the following robustness property.\\

\noindent \textbf{Assumption A} There exist two constants $a>0$, $\theta\in (0,1/2)$, such that, for any  $m_1$ and $m_2\in \mathcal{M}$, there exists a test $\psi_{m_1,m_2}=\psi_{m_2,m_1}$ which chooses between $m_1$ and $m_2$, and satisfies:
\begin{eqnarray}
\sup_{ \set{s \in \mathcal{S} |  d(s,m_1)\leq \theta d(m_1,m_2)}} \mathbb{P}_s\croch{\psi_{m_1,m_2}=m_2}\leq  ~ \exp\paren{-a |\textbf{X}_v| d^{2}(m_1,m_2)}\enspace, \label{Hyp1}
\end{eqnarray}
\begin{eqnarray}
\sup_{ \set{s \in \mathcal{S} |  d(s,m_2)\leq \theta d(m_1,m_2)}} \mathbb{P}_s\croch{\psi_{m_1,m_2}=m_1}\leq  ~ \exp\paren{-a |\textbf{X}_v| d^{2}(m_1,m_2)}\enspace. \label{Hyp2}
\end{eqnarray}

In the density framework Assumption A is fulfilled with $d=h$ for the previous tests (see \cite{Birge1984a,Birge1984b} and \citet[Section 2]{Baraud2011}). In \cite{Birge2013a}, it is shown that $a=(1-2\theta)^2$ for the first test (\ref{test_Birge}). The proof of the robustness when using (\ref{test_Baraud}) in the density framework is an unpublished result of Sart (private communication) leading only to a different value of $a$. Nevertheless, it has been done in a different context by \citet[Section 6]{Sart2011}. 

\section{Efficient algorithms for T-estimation}
In this section, we describe our algorithms which are at the core of the {\it Density.T.HoldOut} package to implement THO. Both algorithms may be useful in a general framework of T-estimation as they allow one to reduce the combinatorial complexity. While our first algorithm computes the true T-estimator, the second implements a lossy approach which reduces the complexity further when the family $S$ is very large, while maintaining good performance in terms of Hellinger risk. In both cases, we assume that {\it Step one} has already been performed, hence our aim is only to select $\myhat m$ among the finite collection $S$ of preliminary estimators using $\textbf{X}_v$, as described in Section \ref{RHO}. Since $\cal M$ is finite, we assume without loss of generality that $\mathcal M=\set{1,\ldots,M}$. Since the estimators $\myhat{s}_{m}[\textbf{X}_t]$ are built from a sample independent of $\textbf{X}_v$, they are, conditionally to $\textbf{X}_t$, deterministic points in $\cal S$. From now on we denote them $s_{m}$ - or $m$ when no confusion is possible - and the THO criterion $\text{crit}_{{\rm THO}}(m,\textbf{X}_t,\textbf{X}_v)$ is denoted $\mathcal{D}(m)=\max_{i \in \mathcal{R}_{m}} d(i,m)$, where we recall that $\mathcal{R}_{m}$ consists of the $j\in\set{1,\ldots,M}\setminus\{m\}$ which are chosen against $m$ by the robust tests. Finally let us denote $\bar{\mathcal{B}}(m,r)=\set{l\in\set{1,\ldots,M}: d(m,l)\leq r }$ the intersection of $\cal M$ with the closed ball with center $m$ and radius $r>0$. From a purely combinatorial point-of-view, the computation of $\myhat m$ minimizing the plausibility index $\mathcal{D}(m)$ requires the computation of $O(M^2)$ tests with a ``naive'' algorithm, which is prohibitive as compared to the $O(M)$ operations needed to compute the classical HO estimator. 

\subsection{Exact T-Hold-Out}\label{sec:algorithm}
The T-estimator search can be realized with a non-quadratic number of tests, thanks to a simple argument which is summarized by the following lemma and its corollary. 

\begin{figure}[H]
\[ \includegraphics[height=7cm,width=6.5cm]{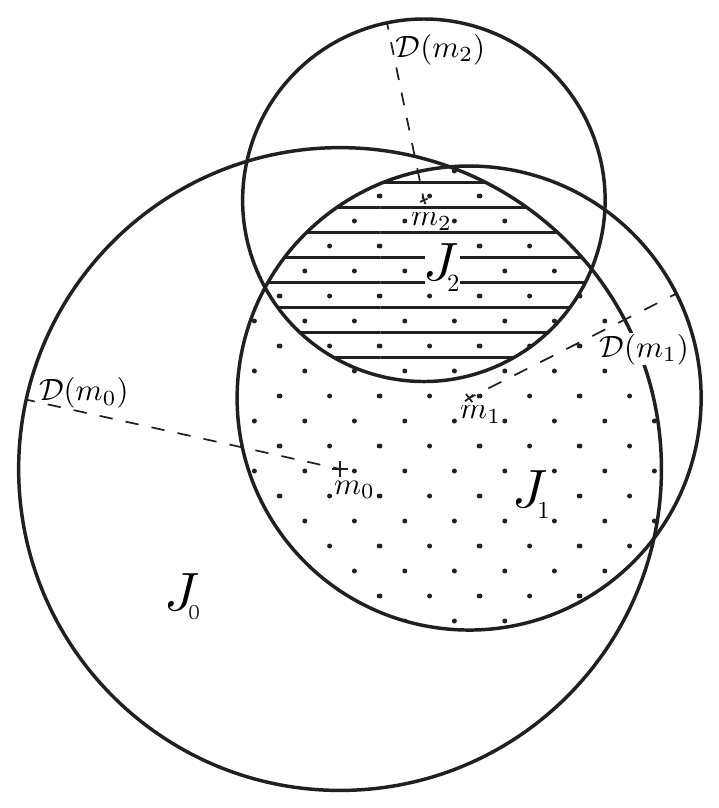} \]
\caption{\label{fig:algo1} \scriptsize Illustration of our exact search for T-estimation. Along the three first iterations, the estimators $m_i$, $i=0,1,2$ are considered with associated radii $\mathcal D(m_i)$ and the T-estimator belongs successively to $J_i$ where $J_0$ is $\bar{\mathcal{B}}(m_0,\mathcal{D}(m_0))$, $J_1$ is the dotted and $J_2$ the hatched area.}
\end{figure}

\begin{Lemma}\label{lem:step 1}
For any point $m_0\in \set{1,\ldots,M}$, the T-estimator $\myhat {m}$ belongs to $\bar{\mathcal{B}}(m_0,\mathcal{D}(m_0))$. 
\end{Lemma}

\begin{proof}
Suppose that there exists one point $m_0\in \set{1,\ldots,M}$ such that $\myhat {m}$ does not belong to the closed ball of radius $\mathcal{D}(m_0)$ centered at $m_0$. Then it does not belong to $\mathcal R_{m_0}$, and it follows that $\psi_{m_0,\myhat m}=m_0$. Hence $m_0$ belongs to $\mathcal R_{\myhat m}$ leading to $\mathcal{D}(\myhat m)\geq d(\myhat m,m_0)> \mathcal{D}(m_0)$ which provides a contradiction with $\mathcal{D}(\myhat m)=\min_{m\in \set{1,\ldots,M}} \mathcal{D}(m)$. 
\end{proof}

\begin{Corollary}\label{lem:step 2}
For any subset $J \subset \set{1,\ldots,M}$, the T-estimator $\myhat {m}$ belongs to 
\[  \bigcap_{m\in J} \bar{\mathcal{B}}(m,\mathcal{D}(m))\enspace. \]
\end{Corollary}

\begin{proof} The proof, illustrated by Figure \ref{fig:algo1}, is straightforward using similar arguments as in Lemma \ref{lem:step 1}.
\end{proof}

It follows that, starting from $m_0$, only a point inside $\bar{\mathcal{B}}(m_0,\mathcal{D}(m_0))$ may be the T-estimator. If any point $m_1$ in this first ball satisfies $\mathcal{D}(m_1)<\mathcal{D}(m_0)$, by Lemma \ref{lem:step 2}, the T-estimator will belong to $\bar{\mathcal{B}}(m_0,\mathcal{D}(m_0))\bigcap \bar{\mathcal{B}}(m_1,\mathcal{D}(m_1))$. Again, criterion $\mathcal{D}$ needs to be computed only for points inside this intersection. We keep intersecting balls $\bar{\mathcal{B}}(m,\mathcal{D}(m))$ until there are no more points with a value of $\mathcal{D}$ smaller than its running value. This approach provides an exact computation of the T-estimator.\\

At each step of the recursion, the current best point is denoted $m$ with associated value $\mathcal{D}(m)$ denoted by $\mathcal{D}$. The running intersection which contains the potentially better points than $m$ is denoted $J$ (this set does not contain $m$). The recursion stops when $J$ is empty. At a given step of the recursion, a point $j$ in $J$ is better than $m$ - and thus replaces it - if $\mathcal{D}(j)<\mathcal{D}$. In all cases, $j$ is removed from the set $J$. During the iteration, $|J|$ and $\mathcal{D}$ decrease ensuring that the algorithm stops. The last running $m$ is the T-estimator. The pseudo-code implementing the efficient and exact search of the T-estimator is provided by Algorithm \ref{algo1}.\\

\begin{algorithm}[H] \footnotesize
\caption{\footnotesize Efficient and exact T-Hold-Out} \label{algo1}
\DontPrintSemicolon
\LinesNumbered
\SetKwInput{KwInit}{Input}
\SetKwInput{KwOut}{Return}
\KwInit{ $m\in J=\set{1,\ldots,M}$ }
\lFor{$(j\neq m)$}{
	compute $\psi_{m,j}(\textbf{X}_v)$
	}
Compute $\mathcal{D}=\mathcal{D}(m)$ and set $J=\bar{\mathcal{B}}(m,\mathcal{D})\setminus \{m\}$ \nllabel{alg:J1}

\While{$(|J|>0)$} {
	Set $\mathcal{D}_{tmp}=0$, select $j\in J$ and set $J=J\setminus \{j\}$ \nllabel{alg:j selection}

	\For{$(k\neq j)$}{ \nllabel{alg:for over k}
		Compute $\psi_{k,j}(\textbf{X}_v)$ \tcp*[f]{if it has not been done yet}
		
		\If(\tcp*[f]{$k\in\mathcal{R}_{j}$}){$(\psi_{k,j}(\textbf{X}_v)==k)$ }{ 	
			Set $\mathcal{D}_{tmp}=\max(\mathcal{D}_{tmp},d(j,k))$
			
			\lIf{$(\mathcal{D}_{tmp}>\mathcal{D})$}{ 
				 break \tcp*[f]{break the for loop}
			 } 
		}
	}				
	Set $m=j$, $\mathcal{D}=\mathcal{D}_{tmp}$ and $J=J\bigcap \bar{\mathcal{B}}(m,\mathcal{D})$ \nllabel{alg:J2}
}
\KwOut{$m$\tcp*[f]{the T-estimator}} 

\end{algorithm}\medskip

\noindent {\it Comments:}  This algorithm works for all the statistical frameworks of T-estimation, and does not depend on the considered robust test. The ``for'' loop is realized on all $k\not=j$, as $\mathcal{D}(k)$ depends on all points and not only on those in $J$. If there are $N$ points in the first ball, the number of computed tests is at most $O(N*M)$.  Moreover, if the first ball is empty, i.e. if $\mathcal{D}(m)=0$, the algorithm stops immediately, returning $m$ for $\myhat m$. In this case, the complexity of our algorithm is $O(M)$. Any preliminary estimator (maximum likelihood, least-squares, $L_1$-minimizer, etc.) may be a starting point of our algorithm. We hope that by beginning from a good preliminary estimator, there will be only few points in the first ball, resulting in less computations.  The computation requires $O(M^2)$ operations if  $J$ decreases by only one point at each step of the recursion which happens only if the selected $j$ satisfies
	\[ \max_{k\in J}d(j,k) = \max_{k\neq l\in J} d(k,l)\]
	at each iteration.\medskip

\subsection{Fast algorithm for approximate T-Hold-Out}
Assumption A ensures that as soon as the Hellinger distance between two estimators of $S$ is large enough, the probability that the robust test does not choose the best estimator is small. However, as shown in Lemma 1 of \citet{LeCam1973}, when this distance is smaller than $c\,n^{-1/2}$, where $c$ is a small positive constant, the two corresponding probabilities cannot be separated by a test built on $n$ observations anymore. From this remark, we derive a lossy version from our efficient and exact algorithm. The main difference consists in ignoring points in $S$ as soon as their Hellinger distance to a previously considered one is smaller than a given threshold $\delta_n>0$. 

\begin{figure}[H]
\[ \includegraphics[height=7cm,width=6.5cm]{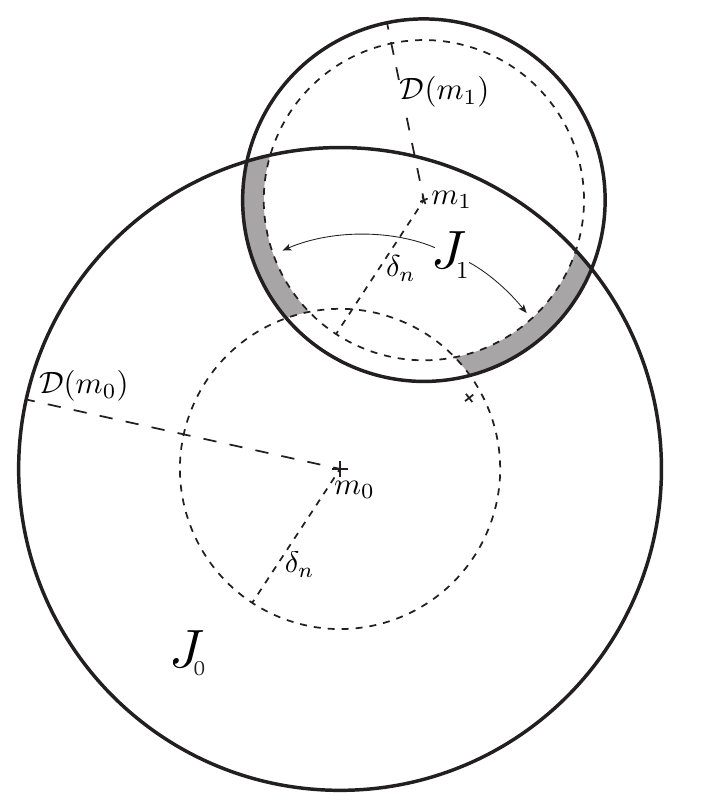} \]
\caption{\label{fig:algo2} \scriptsize Illustration of the approximate T-estimation search: $J_0$ is a ring around $m_0$. The point following $m_0$ has changed with respect to Figure \ref{fig:algo1} as the previously selected $m_1$ is now inside $\bar{\mathcal{B}}(m_0,\delta_n)$. $J_1$ (in grey) appears as the intersection of two rings.}
\end{figure}

We introduce this distance control at two steps of our efficient and exact algorithm. As the interior points of $\bar{\mathcal{B}}(m,\delta_n)$ cannot be properly distinguished from $m$ by any test, the set $J$ becomes, at lines \ref{alg:J1} and \ref{alg:J2} of Algorithm \ref{algo1}, the intersection of rings instead of balls, obtained by removing from the original ball $\bar{\mathcal{B}}(m,\mathcal{D}(m))$ the ball $\bar{\mathcal{B}}(m,\delta_n)$. In the same spirit, at line \ref{alg:for over k} of Algorithm \ref{algo1}, the current $k$, in the {\it for} loop, is considered if and only if its distance to $\mathcal T_j$ is larger than $\delta_n$, where $\mathcal T_j$ is made of the running $j$ and the further points which have been tested against $j$. The pseudo-code of this lossy version is provided by Algorithm \ref{algo2} and illustrated by Figure \ref{fig:algo2}.

\begin{algorithm}[H] \footnotesize
\caption{\footnotesize Approximate T-Hold-Out} \label{algo2}
\DontPrintSemicolon
\LinesNumbered
\SetKwInput{KwInit}{Input}
\SetKwInput{KwOut}{Return}
\KwInit{ $m\in J=\set{1,\ldots,M}$; $\delta_n>0$}

\lFor{$(j\neq m)$}{
	compute $\psi_{m,j}(\textbf{X}_v)$
	}
Compute $\mathcal{D}=\mathcal{D}(m)$ and set $J=\bar{\mathcal{B}}(m,\mathcal{D})\setminus\bar{\mathcal{B}}(m,\delta_n)$

\While{$(|J|>0)$} {
	Set $\mathcal{D}_{tmp}=0$, select $j\in J$ and set $J=J\setminus \{j\}$ \nllabel{alg2:j selection}
	
	Define $\mathcal T_j=\{j\}$

	\For{$(k\neq j)$}{
	
		\lIf{$(d(j,\mathcal T_j)\leq \delta_n)$} {next $k$ \tcp*[f]{next $k$ if distance is too small}} 
		
		Set $\mathcal T_j=\mathcal T_j\cup\{k\}$
		
		Compute $\psi_{k,j}(\textbf{X}_v)$ \tcp*[f]{if it has not been done yet}
		
		\If(\tcp*[f]{$k\in\mathcal{R}_{j}$}){$(\psi_{k,j}(\textbf{X}_v)==k)$ }{ 	
			Set $\mathcal{D}_{tmp}=\max(\mathcal{D}_{tmp},d(j,k))$
			
			\lIf{$(\mathcal{D}_{tmp}>\mathcal{D})$}{ 
				 break \tcp*[f]{break the for loop}
			 } 
		}
	}				
	Set $m=j$, $\mathcal{D}=\mathcal{D}_{tmp}$ and $J=J\bigcap \left[\bar{\mathcal{B}}(m,\mathcal{D})\setminus\bar{\mathcal{B}}(m,\delta_n)\right]$
}
\KwOut{$m$\tcp*[f]{the approximate T-estimator}}

\end{algorithm}

\section{Simulation protocol}\label{sec:protocol}

 In our simulations, we consider only the density estimation framework. This is motivated by the fact that likelihood ratio tests are not robust in this context, and we hoped to observe differences in terms of risk. \smallskip
 
We considered $\textbf{X}=\set{X_{1},\ldots,X_{n}}$ i.i.d. random variables from an unknown density $s$ with respect to the Lebesgue measure on $\mathcal{X}=\mathbb{R}$ and, for a given proportion $p$ in $(0,1)$, we divide randomly $\textbf{X}$ into $\textbf{X}_t=\set{X_1,\ldots,X_{n_1}}$ and $\textbf{X}_v=\set{X_{n_1+1},\ldots,X_n}$, with $n_1=[pn]$ where $[x]$ is the integer part of $x$.
Simulations were carried out with four sample sizes $n=100,250,500,1000$ and three different proportions $p=1/2, 2/3, 3/4$ using the two different robust tests (\ref{test_Birge}) and (\ref{test_Baraud}). Our test functions $s$ vary in a subset $\mathcal{L}$ made of the densities $s_1$,\ldots, $s_{28}$ of the R-package {\it benchden}\footnote{{\it Benchden} (see \cite{Mildenberger-Weinert2012}) implements the benchmark distributions of \citet{Berlinet-Devroye1994}. Available on the CRAN \url{http://cran.r-project.org/web/packages/benchden/index.html}.} which are in $L_1\cap L_2$  - to ensure that risks are computable. This set $\mathcal{L}$ is made of the densities $s_i$ for $$i\in\set{1,\ldots,5,7,11,12,13,16,17,21,\ldots,27}.$$
We considered several estimator collections: 
\begin{itemize}
\item $S_R$ made of regular histograms with bin number varying from 1 to $\lceil n_1/\log(n_1)\rceil$ as described in \citet{Birge-Rozenholc2006};
\item $S_I$ made of the maximum likelihood irregular histograms when the bin number only varies from 1 to $\min(100,\lceil n_1/\log(n_1)\rceil)$ as described in \citet{Rozenholc-Mildenberger-Gather2010};
\item $S_K$ made of Gaussian kernel estimators with the varying bandwidths chosen as $$(\max[\textbf{X}_t]-\min[\textbf{X}_t])/2j\quad\mbox{ for }\quad j=1,\ldots,\lceil n_1/\log(n_1) \rceil.$$
\item $S_P$ made of parametric estimates obtained by moment's method for the Gaussian, exponential, log-normal, chi-square, gamma and beta distributions together with a maximum likelihood estimate of the uniform distribution;
\item $S_C=S_R \cup S_I $ 
\item $S_1=S_R\cup S_I\cup S_K$;
\item $S_2=S_R \cup S_I \cup S_K \cup S_P$.
\end{itemize}
The estimation accuracy of a given procedure $\mytilde s$ has been evaluated using an empirical version of the risk $R_s(\mytilde{s},\ell)=\mathbb{E}_s[\ell(\mytilde{s},s)]$, obtained by generating $100$ $n$-samples $\textbf{X}^{(j)}$, $1\leq j \leq 100$, of density $s$:
\begin{equation*}
\bar{R}_s(\mytilde s,\ell)=\frac{1}{100} \sum_{j=1}^{100} \ell(\mytilde{s}[\textbf{X}^{(j)}],s)\enspace,
\end{equation*}
where $\ell(t,u)$ is either $h^2(t,u)$ or $\|t-u\|_q^q$, for $q=1,2$.\\

In order to compare two  procedures $\mytilde t_1$ and $\mytilde t_2$, we introduce the normalized $\log_2$-ratio of their empirical risks, namely:
\begin{equation*}
\bar{W}_s(\mytilde t_1,\mytilde t_2)= \frac 1 r \log_2\frac{\bar{R}_s(\mytilde t_1,\ell)}{\bar{R}_s(\mytilde t_2,\ell)} = \log_2\bar{R}_s^{1/r}(\mytilde t_1,\ell) - \log_2\bar{R}_s^{1/r}(\mytilde t_2,\ell)\enspace,
\end{equation*}
where $r$ is equal to $q$ for $L_q$ losses and 2 for the Hellinger loss. The aim of the normalization by $r$ is to provide an easier comparison of $\bar W_s$ when the loss changes. In our empirical study, procedure $\mytilde t_2$ is thus considered better in terms of risk than $\mytilde t_1$ for a given loss function $\ell$ if the values of $\bar{W}_s(\mytilde t_1,\mytilde t_2)$ are positive when the density $s$ varies.\\

We compared the four hold-out methods described above: T-estimation with the tests given by (\ref{test_Birge}) and (\ref{test_Baraud}), LS and KL. We first computed $\myhat s_m[\textbf{X}_t]$ for all $m\in \cal M$, and then selected $\hat m$ minimizing the respective HO criterion resulting in $\myhat m_{T1}$, $\myhat m_{T2}$, $\myhat m_{LS}$ and $\myhat m_{KL}$, providing $\mytilde s$ as either $\myhat s_{\myhat m}[\textbf{X}_t]$ or $\myhat s_{\myhat m}[\textbf{X}]$. As $\myhat m$ depends on the chosen proportion $p$, in order to explicitly specify the dependency of $\myhat m$ with respect to this parameter, we will use the following notations $\myhat s_{\myhat m[p]}[\textbf{X}_t]$ or $\myhat s_{\myhat m[p]}[\textbf{X}]$ when needed. In Algorithms \ref{algo1} and \ref{algo2}, the input $m$ has been set to $\myhat m_{LS}$  and $j=\arg\max_{k\in J} d(k,m)$, at line \ref{alg:j selection}. In Algorithm \ref{algo2}, we fixed $\delta_n=1/\sqrt{ |\textbf{X}_v| }$ as a lower bound for the Hellinger distance between distinguishable probabilities, following \citet{LeCam1973}.\\

Moreover, we also considered some calibrated estimation procedures which choose $m$ in some particular families. These are not direct competitors with the T-estimation as they cannot deal with general families $S$ but provide a good benchmark in terms of risk: 
\begin{itemize}
\item for $S_R$, $S_I$, $S_C$, the penalized maximum likelihood estimators, denoted $\mytilde s_{\rm pen}$ introduced in \citet{Birge-Rozenholc2006,Rozenholc-Mildenberger-Gather2010} and implemented in the R-package\footnote{available on the CRAN \url{http://cran.r-project.org/web/packages/histogram/index.html}.\label{foot:cran}} {\it histogram},
\item for $S_K$, the ${L}_1$-version of the procedure introduced in \citet{Goldenshluger-Lepski2011}, denoted $\mytilde s_{GL}$. \end{itemize}
For fairness, we applied these calibrated estimation procedures in their original setting which use the full sample replacing $n_1$ by $n$ in the definition of $S_R$ and $S_K$. 

Finally, for the family $S_K$, we considered some  bandwidth selectors (namely nrd, ucv, bcv, SJ) implemented in the {\it density} generic function available in R , providing some well-known estimators $\mytilde s_{nrd}$, $\mytilde s_{bcv}$, $\mytilde s_{ucv}$, $\mytilde s_{SJ}$ of the density which are not chosen in $S$ \citep{Silverman1986,Sheather-Jones1991,Scott1992}. \\

The R-package\footnote{available on the CRAN \url{http://cran.r-project.org/web/packages}} {\it Density.T.HoldOut} is a ready-to-use software that implements our algorithms in the density framework. The main function  - called {\tt DensityTestim} -  receives as input a sample $\textbf{X}$ and a family of estimators and returns the selected estimator. The previously described families are available and can be extended or adapted by the user (default family is $S_2$). Other important input arguments are parameters  $p$, $\theta$  and the starting point  (default values are $p=1/2$ , $\theta=1/4$ and $\myhat m_{LS}$). This function implements the exact and lossy algorithms, through the numeric {\tt csqrt} (default value 1) which controls $\delta_n={\tt csqrt}/\sqrt{|\textbf{X}_v|}$ in Algorithm \ref{algo2}. The robust test might be the one defined by (\ref{test_Birge})  setting {\tt test}='birge' (default), or by (\ref{test_Baraud}) setting {\tt test}='baraud'. The resulting estimator is either built with  $\textbf{X}_t$ ({\tt last}='training') or $\textbf{X}$ ({\tt last}='full', default). 

\section{Simulation results}

This section, made using Algorithm \ref{algo1}, is devoted to the study of the quality of the T-hold-out. We illustrate our results showing boxplots of $\bar W_s(\mytilde t_1,\mytilde t_2)$ for all 18 densities $s\in \mathcal{L}$, various choices of estimators $\mytilde t_1$ and $\mytilde t_2$ and for different collections of estimators $S$, as described in the previous section. We begin by investigating how parameter $\theta$  influences the THO procedure deduced from (\ref{test_Birge}). Then we show that the two robust procedures derived from (\ref{test_Birge}) and (\ref{test_Baraud}) have similar behavior in terms of risk, and therefore pursue using the first one only. After studying how $p$ influences the quality of estimation, we provide two main comparison types. First we look at HO methods which select among a family of points using the validation sample. Then we compare the THO against some density estimation methods, which are not necessarily selection procedures anymore. In this subsection, we divide the presentation between calibrated selection procedures build directly on the full sample and some selectors of the bandwidth obtained using asymptotic derivation of the risk for some specific loss. 

\subsection{Influence of $\theta$}\label{Theta}
The robustness of the procedure build using (\ref{test_Birge}) is controlled through the parameter $\theta<1/2$ (see Eq. \ref{test_Birge}), the KLHO corresponding to $\theta=0$ (no robustness). We computed the empirical risk using the THO procedure with $\theta=1/16, 1/8, 1/4, 3/8, 7/16$, and $n=100, 250, 500, 1000$. We observed that $\theta$ has little influence in terms of risk ($\theta=1/16$ being slightly worse) and decided to pursue the empirical study with $\theta=1/4$.\smallskip

\subsection{Influence of the robust test}\label{sec: robust test}

\begin{figure}[H]
\[ \includegraphics[height=2.5cm,width=8.5cm]{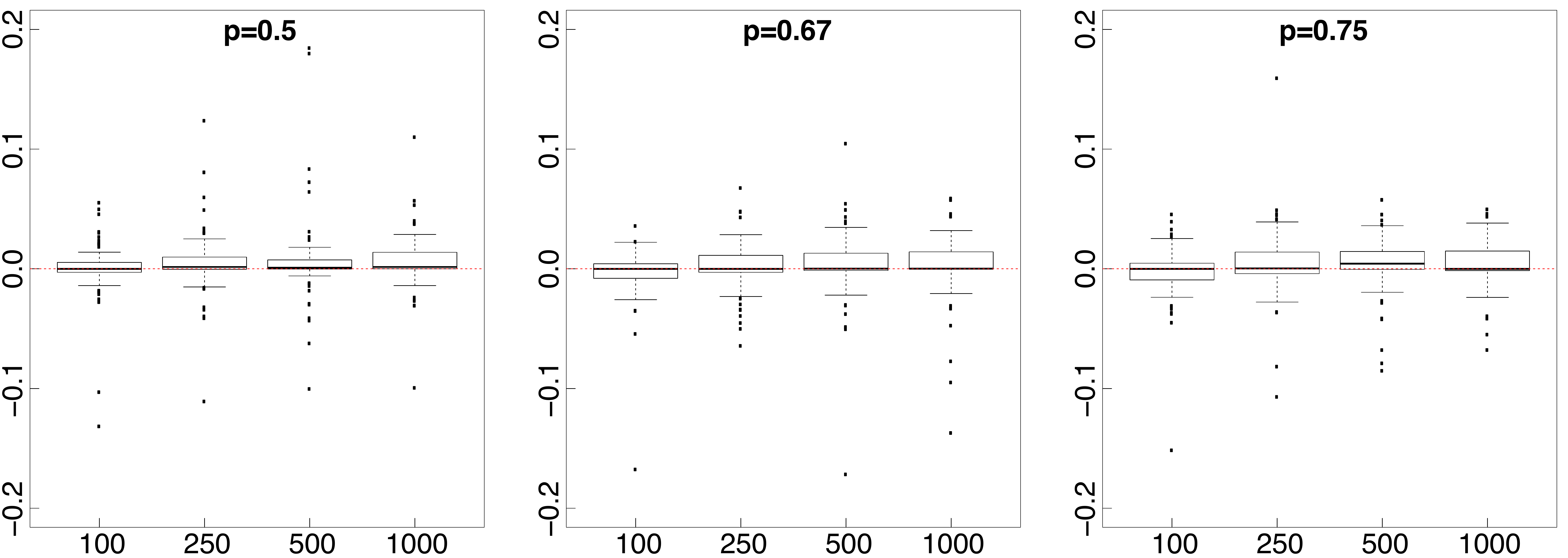} \]
\caption{\label{fig:robust test comparison} \scriptsize From left to right, normalized $\log_2$-ratio of the empirical risks $\bar{W}_s(\myhat s_{\myhat m_{T1}[p]}[\textbf{X}_t],\myhat s_{\myhat m_{T2}[p]}[\textbf{X}_t])$ for the Hellinger loss for $p=1/2$, $2/3$ and $3/4$. Each subfigure shows the boxplot for $n$ equals 100, 250, 500 and 1000. The horizontal red dotted line provides the reference value 0.}
\end{figure}

As we dispose of two robust tests to proceed the THO, we compare the two corresponding strategies in Figure \ref{fig:robust test comparison} using $\mytilde t_1=\myhat s_{\myhat m_{T1}[p]}[\textbf{X}_t]$ and $\mytilde t_2=\myhat s_{\myhat m_{T2}[p]}[\textbf{X}_t]$ for $p=1/2$, $2/3$ and $3/4$ (each value corresponding to one subfigure below). For a fixed $n$, there are $18\times 6$ ratios obtained when both the density and the collection of estimators vary.

Surprisingly the two procedures behave very similarly in all settings, and only few differences can be observed in terms of Hellinger risk (generally less than 2\%). We therefore pursue our empirical study with the procedure derived from (\ref{test_Birge}), and from now on we denote $\hat m_T$ instead of $\hat m_{T1}$, when no confusion is possible.

\subsection{Influence of $p$}\label{sec: p influence}

We examine the dependence of the THO with respect to $p$, the proportion of the initial sample dedicated to building the estimators, using the Hellinger risk. 

\begin{figure}[H]
\[ \includegraphics[height=2.5cm,width=\textwidth]{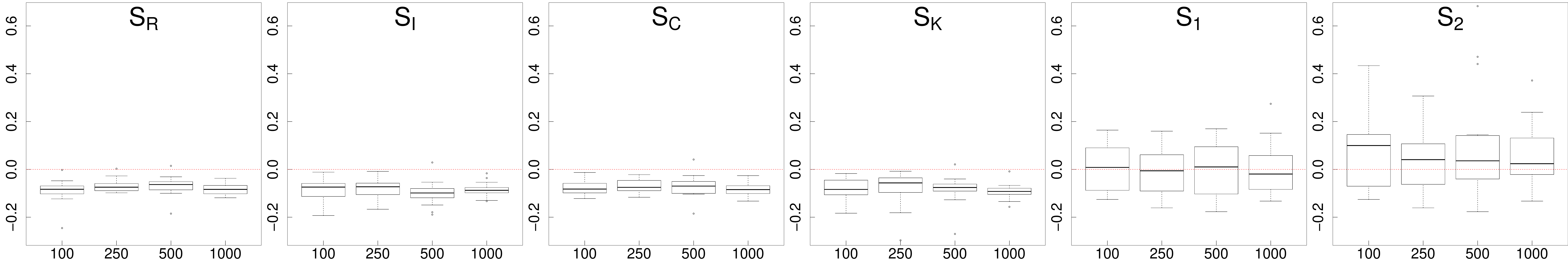} \]
\[ \includegraphics[height=2.5cm,width=\textwidth]{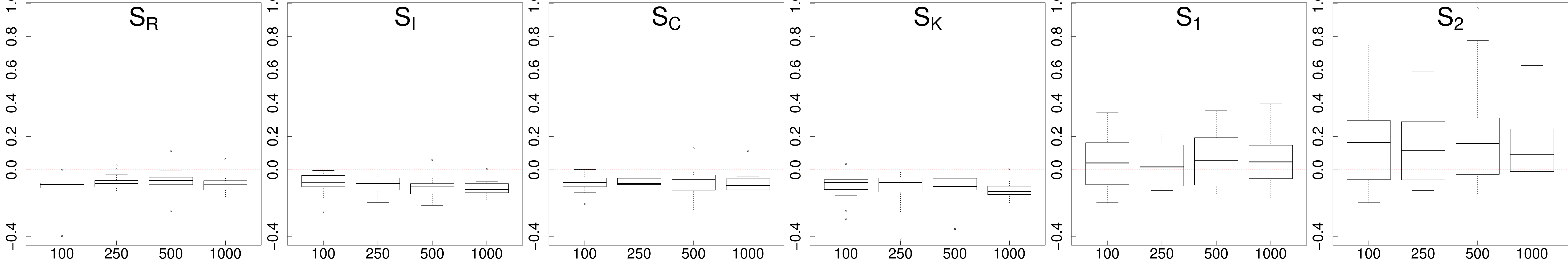} \]
\caption{\label{fig:p in T1} \scriptsize From left to right, normalized $\log_2$-ratio of the empirical risks $\bar{W}_s(\myhat s_{\myhat m_T[2/3]}[\textbf{X}_t],\myhat s_{\myhat m_T[1/2]}[\textbf{X}_t])$ (upper line) and $\bar{W}_s(\myhat s_{\myhat m_T[3/4]}[\textbf{X}_t],\myhat s_{\myhat m_T[1/2]}[\textbf{X}_t])$ (bottom line) for the Hellinger loss, using collections $S_R$, $S_I$, $S_C$, $S_K$, $S_1$ and $S_2$. Each subfigure shows the boxplot for $n$ equals 100, 250, 500 and 1000. The horizontal red dotted line provides the reference value 0.}
\end{figure}

Figure \ref{fig:p in T1} is built using $\mytilde t_1=\myhat s_{\myhat m_T[p]}[\textbf{X}_t]$ for $p$ equals 2/3 (upper line), 3/4 (bottom line) and $\mytilde t_2=\myhat s_{\myhat m_T[1/2]}[\textbf{X}_t]$. We observe two different behaviors for families $S_R$, $S_I$, $S_C$ and $S_K$ on the one hand and for $S_1$ and $S_2$ on the other hand. For the first families $p=2/3$ or $3/4$ is  better than $p=1/2$. For the second ones $p=2/3$ seems equivalent to $p=1/2$ but $p=3/4$ is worst than $p=1/2$. Hence we consider preferable to use $p=2/3$, which makes the best compromise for all families.

\subsection{Comparing Hold-Out methods}

Hold-out procedures are universal since they do not depend on the choice of family $S$. They can be seen as methods that choose among some family of fixed points. Setting $p=2/3$, we compare the THO to the KLHO and LSHO introduced in Section \ref{sec: HO} using each of the 6 estimator collections described in Section \ref{sec:protocol}.

\begin{figure}[H]
\[ \includegraphics[height=2.5cm,width=\textwidth]{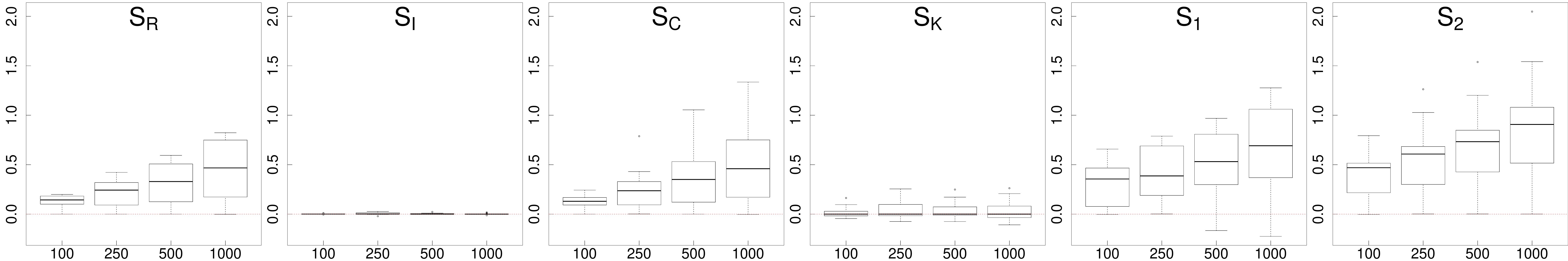} \]
\[ \includegraphics[height=2.5cm,width=\textwidth]{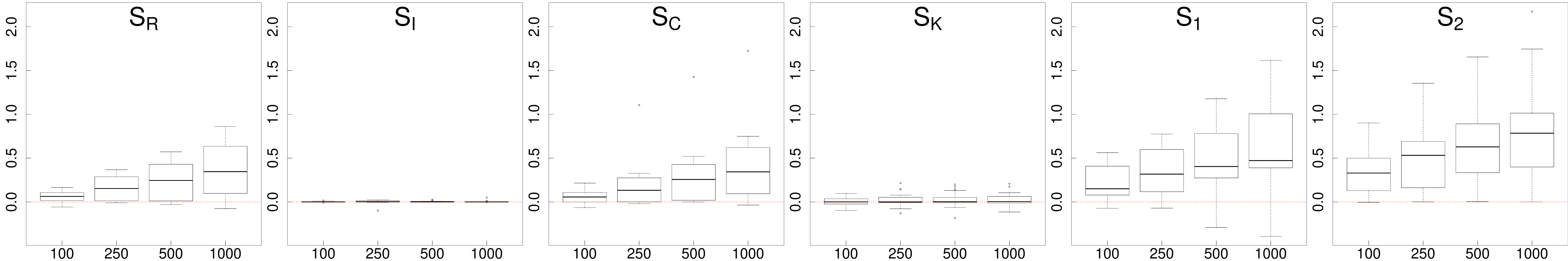} \]
\caption{\label{fig:KL vs T1} \scriptsize From left to right, normalized $\log_2$-ratio of the empirical risks  $\bar{W}_s(\myhat s_{\myhat m_{KL}}[\textbf{X}_t],\myhat s_{\myhat m_T}[\textbf{X}_t])$ for $p=2/3$, using collections $S_R$, $S_I$, $S_C$, $S_K$, $S_1$ and $S_2$. Upper line, using Hellinger loss, bottom line using $L_1$ loss. See Figure \ref{fig:p in T1} for more details.}
\end{figure}
Figure \ref{fig:KL vs T1} is built using $\mytilde t_1=\myhat s_{\myhat m_{KL}}[\textbf{X}_t]$ and $\mytilde t_2=\myhat s_{\myhat m_T}[\textbf{X}_t]$ considering Hellinger (upper line) and $L_1$  (bottom line) losses. In all cases, the median and most of the distribution are positive, meaning that the THO outperforms the KLHO estimator. For collections $S_I$ and $S_K$, empirical risks for both losses are similar, with $\bar{W}_s(\myhat s_{\myhat m_{KL}}[\textbf{X}_t],\myhat s_{\myhat m_T}[\textbf{X}_t])$ being respectively larger than -0.01 (except for the uniform density) for $S_I$, and -0.2 for $S_K$.  When $n$ grows, while for $S_I$ and $S_K$ the ratio remains stable, it increases for all other families in favor of the THO. Moreover when going from collection $S_1$ to $S_2$, that is adding the parametric collection $S_P$, we observe that the already good performance of the THO improves. We therefore suspect that the THO chooses the parametric estimator more often than KLHO when facing the corresponding densities. 

\begin{figure}[H]
\[ \includegraphics[height=2.5cm,width=\textwidth]{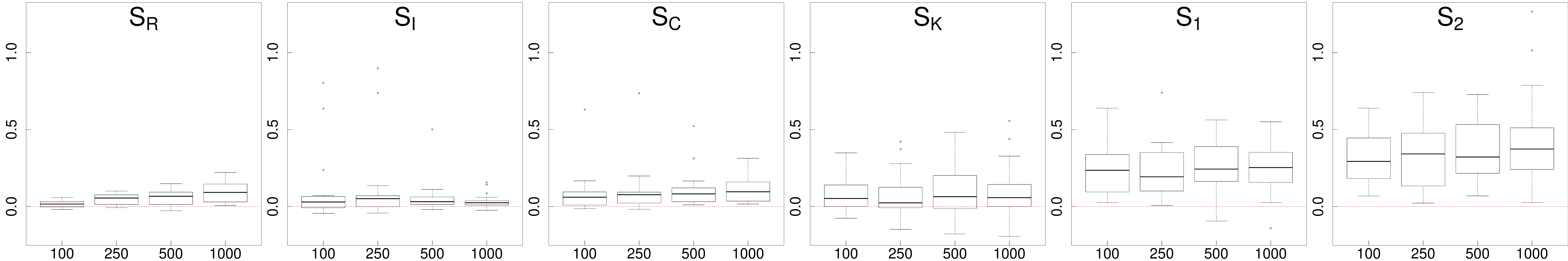} \]
\[ \includegraphics[height=2.5cm,width=\textwidth]{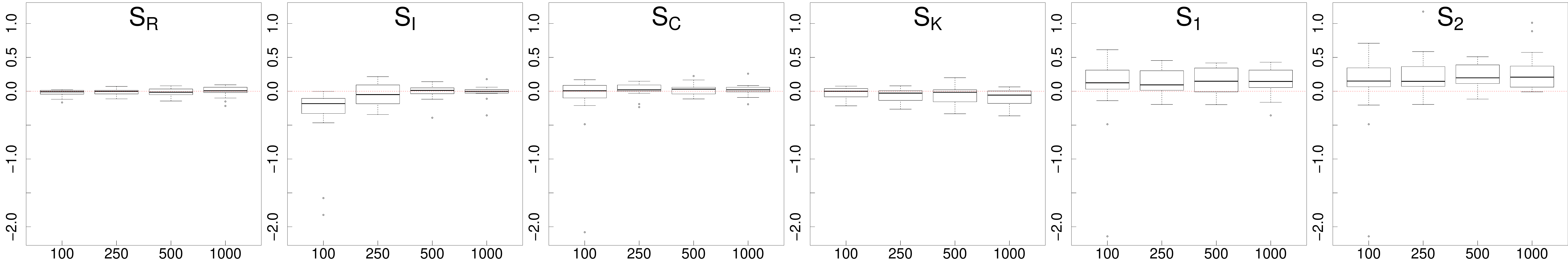} \]
\caption{\label{fig:LS vs T1} \scriptsize From left to right, normalized $\log_2$-ratio of the empirical risks $\bar{W}_s(\myhat s_{\myhat m_{LS}}[\textbf{X}_t],\myhat s_{\myhat m_T}[\textbf{X}_t])$ for $p=2/3$, using collections $S_R$, $S_I$, $S_C$, $S_K$, $S_1$ and $S_2$. Upper line, using Hellinger loss, bottom line using $L_2$ loss. See Figure \ref{fig:p in T1} for more details.}
\end{figure}
Figure \ref{fig:LS vs T1} is built using $\mytilde t_1=\myhat s_{\myhat m_{LS}}[\textbf{X}_t]$ and $\mytilde t_2=\myhat s_{\myhat m_T}[\textbf{X}_t]$ considering Hellinger (upper line) and $L_2$  (bottom line) losses. The THO performs better than the LSHO estimator for all collections except for the collection $S_I$ when $n=100$. For the larger collections $S_1$ and $S_2$, the THO outperforms the LSHO. However, as $n$ grows, we observe that the relative quality of the two procedures remain stable.

\subsection{Comparing final strategies for T-Hold-Out}

\begin{figure}[H]
\[ \includegraphics[height=2.5cm,width=\textwidth]{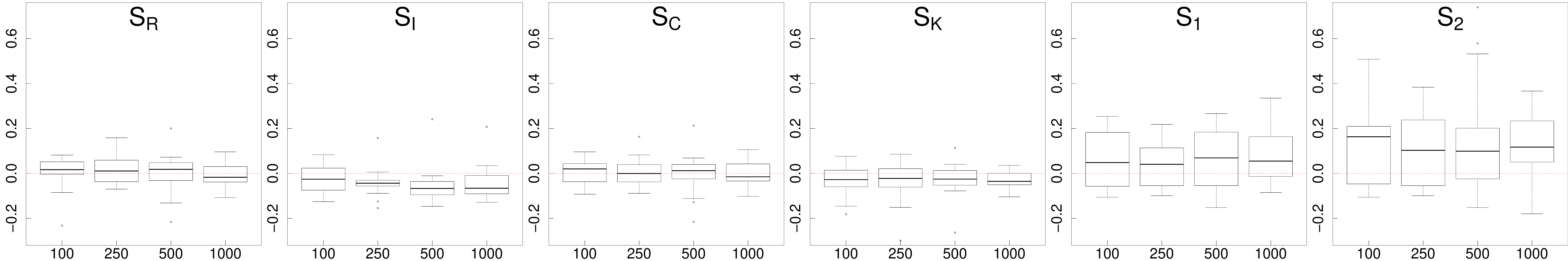} \]
\[ \includegraphics[height=2.5cm,width=\textwidth]{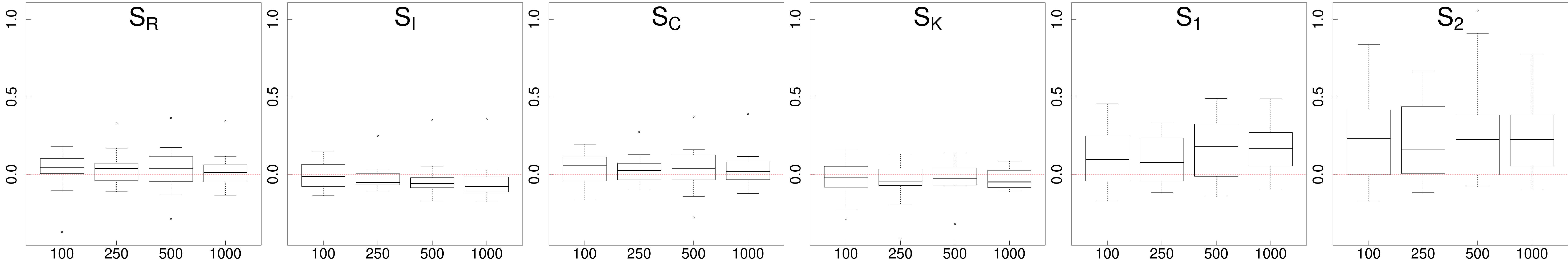} \]
\caption{\label{fig:p in T2} \scriptsize From left to right, normalized $\log_2$-ratio of the empirical risks $\bar{W}_s(\myhat s_{\myhat m_T[2/3]}[\textbf{X}],\myhat s_{\myhat m_T[1/2]}[\textbf{X}])$ (upper line) and $\bar{W}_s(\myhat s_{\myhat m_T[3/4]}[\textbf{X}],\myhat s_{\myhat m_T[1/2]}[\textbf{X}])$ (bottom line) for the Hellinger loss, using collections $S_R$, $S_I$, $S_C$, $S_K$, $S_1$ and $S_2$. See Figure \ref{fig:p in T1} for more details.}
\end{figure}

Here, we investigate whether $\myhat s_{\myhat m_T}[\textbf{X}_t]$ or $\myhat s_{\myhat m_T}[\textbf{X}]$ performs better. For this purpose, we study the Hellinger risk of $\myhat s_{\myhat m_T}[\textbf{X}]$ when $p$ varies. Figure \ref{fig:p in T2} is built using $\mytilde t_1=\myhat s_{\myhat m_T[p]}[\textbf{X}]$ for $p$ equals 2/3 (upper line), 3/4 (bottom line) and $\mytilde t_2=\myhat s_{\myhat m_T[1/2]}[\textbf{X}]$. We observe that against $p=2/3$ or $p=3/4$, the value $p=1/2$ provides better results for the large families $S_1$ and $S_2$ while for the small families the results are more balanced. Hence we consider preferable to make use of this strategy with $p=1/2$. 

\begin{figure}[H]
\[ \includegraphics[height=2.5cm,width=\textwidth]{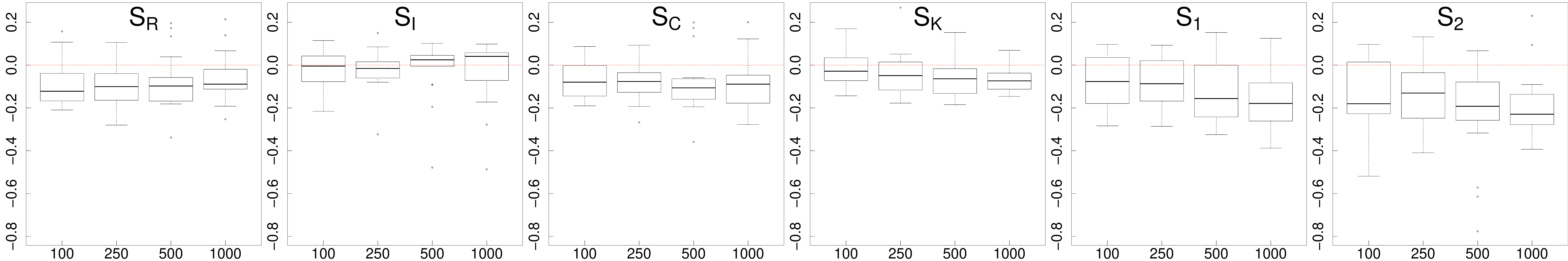} \]
\caption{\label{fig:T1 vs T2} \scriptsize From left to right, normalized $\log_2$-ratio of the empirical risks $\bar{W}_s(\myhat s_{\myhat m_T[1/2]}[\textbf{X}],\myhat s_{\myhat m_T[2/3]}[\textbf{X}_t])$ for Hellinger loss, using collections $S_R$, $S_I$, $S_C$, $S_K$, $S_1$ and $S_2$. See Figure \ref{fig:p in T1} for more details.}
\end{figure}

We now compare the Hellinger risks of $\myhat s_{\myhat m_T[2/3]}[\textbf{X}_t]$  - which appeared as the best competitor in Section \ref{sec: p influence} -  and $\myhat s_{\myhat m_T[1/2]}[\textbf{X}]$. Figure \ref{fig:T1 vs T2} is built using $\mytilde t_1=\myhat s_{\myhat m_T[1/2]}[\textbf{X}]$ and $\mytilde t_2=\myhat s_{\myhat m_T[2/3]}[\textbf{X}_t]$. We observe that the strategy $\myhat s_{\myhat m_T[1/2]}[\textbf{X}]$ is preferable, since its median (and even most of its distribution) is negative in all considered settings. It should be noticed that our simulations show that, more than the value of $p$, it is the use of $\textbf{X}$ instead of $\textbf{X}_t$ which has the larger influence on the final risk.


\subsection{T-Hold-Out against dedicated estimation procedures}
\begin{figure}[H]
\[ \includegraphics[height=2.5cm,width=0.7\textwidth]{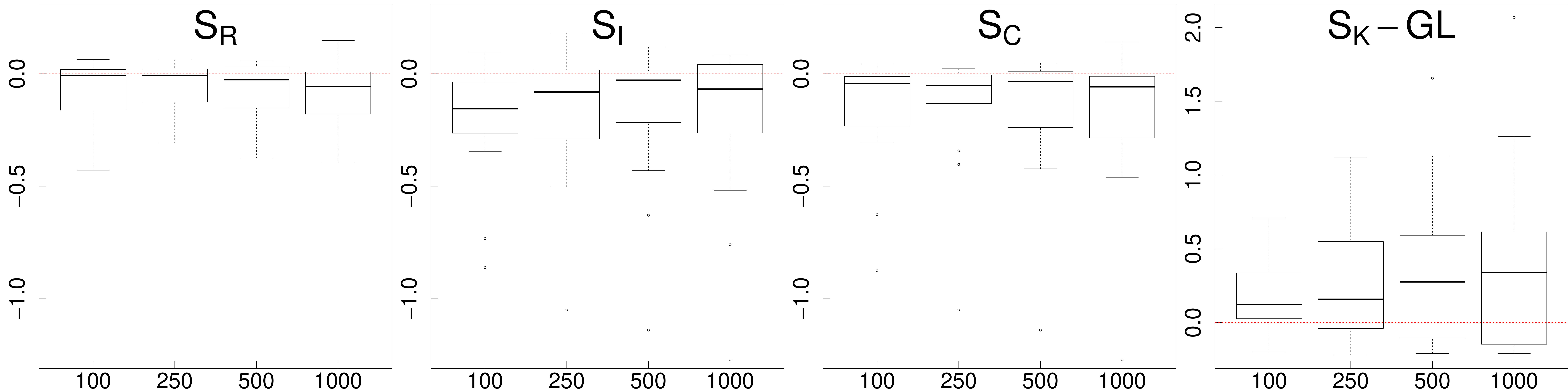} \]
\[ \includegraphics[height=2.5cm,width=0.7\textwidth]{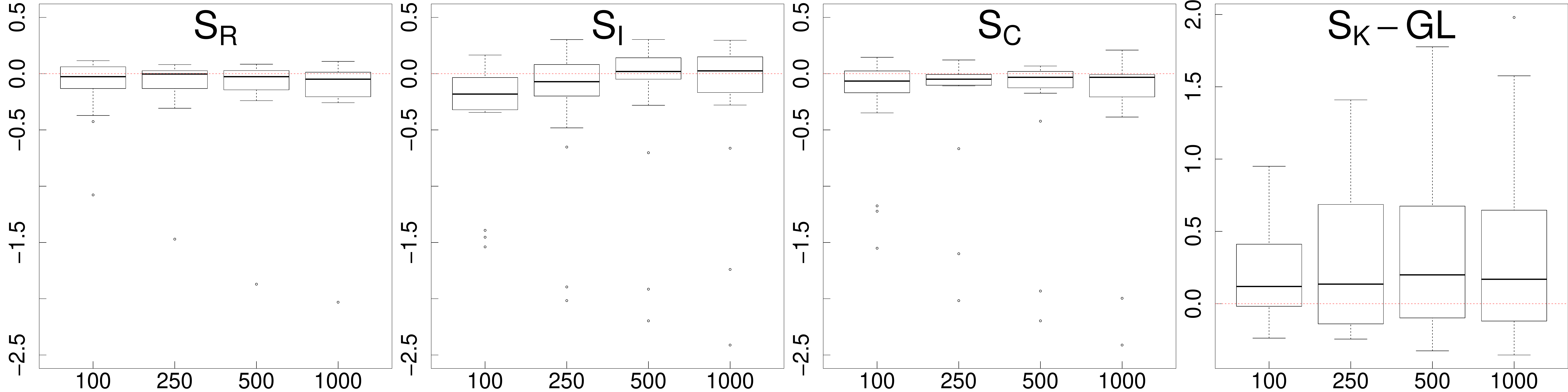} \]
\caption{\label{fig:best vs T2} \scriptsize From left to right, normalized $\log_2$-ratio of the empirical risks $\bar{W}_s(\mytilde s[\textbf{X}],\myhat s_{\myhat m_T[1/2]}[\textbf{X}])$ using collections $S_R$, $S_I$, $S_C$ and $S_K$ with Hellinger (upper line) and $L_1$ (bottom line) losses.  For the 3 first collections $\mytilde s$ is $\mytilde s_{\rm pen}$ and $\mytilde s_{GL}$ for $S_K$. Each subfigure shows the boxplot for $n$ equals 100, 250, 500 and 1000. The horizontal red dotted line provides the reference value 0.}
\end{figure}

We now compare the THO competitor $\myhat s_{\myhat m_T[1/2]}[\textbf{X}]$ against the so-called dedicated methods. Figure \ref{fig:best vs T2} is built using $\mytilde t_1=\mytilde s[\textbf{X}]$ ($\mytilde s$ being either $\mytilde s_{\rm pen}$ or $\mytilde s_{GL}$) and $\mytilde t_2=\myhat s_{\myhat m_T[1/2]}[\textbf{X}]$ considering Hellinger (upper line) and $L_1$ (bottom line) losses. We observe that the THO is slightly worse than a well-calibrated procedure for  histograms but outperforms the $L_1$-version of the Goldenshluger-Lepski procedure. 

\begin{figure}[H]
\[ \includegraphics[height=2.5cm,width=0.7\textwidth]{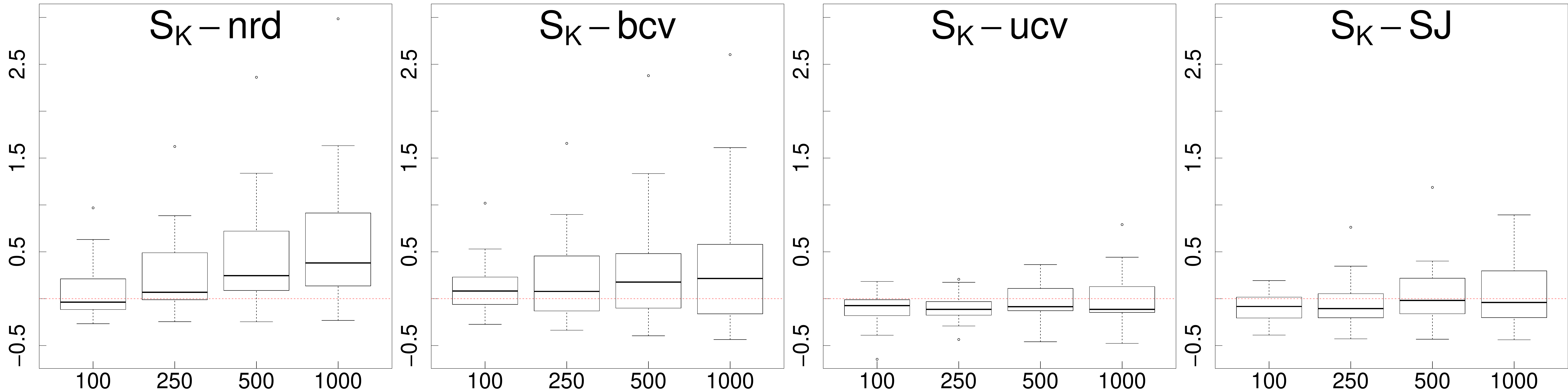} \]
\[ \includegraphics[height=2.5cm,width=0.7\textwidth]{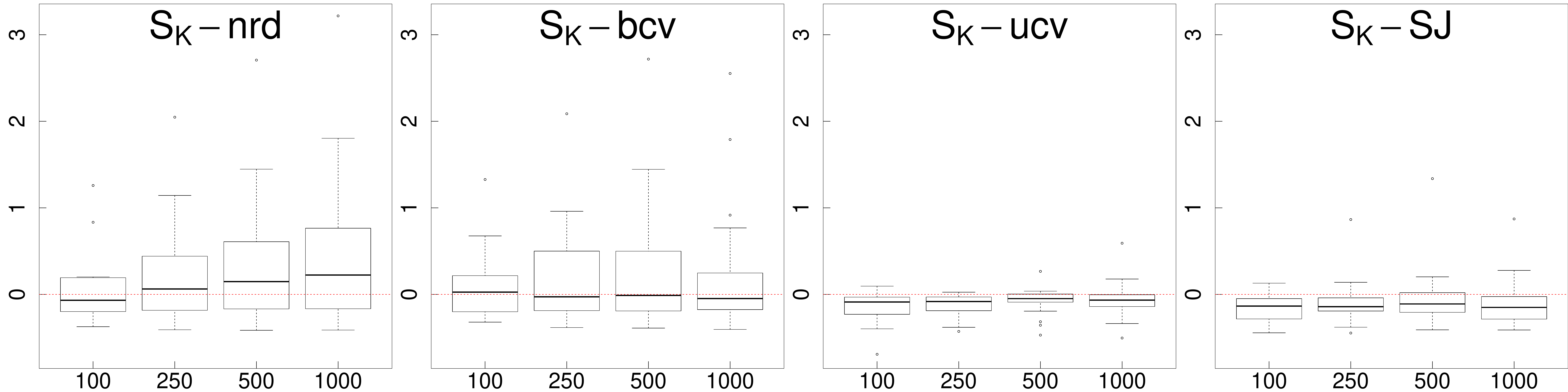} \]
\caption{\label{fig:other vs T2} \scriptsize From left to right, normalized $\log_2$-ratio of the empirical risks $\bar{W}_s(\mytilde s[\textbf{X}],\myhat s_{\myhat m_T[1/2]}[\textbf{X}])$ for collection $S_K$. The 3 first competitors $\mytilde s$ are the kernel estimators with respective bandwidth provided by the bandwidth selectors $nrd$, $bcv$, $ucv$ and $SJ$ as defined in the function {\it density} of the {\it stats} package of {\it R}. Upper line, using Hellinger loss, bottom line using $L_1$ loss. See Figure \ref{fig:p in T1} for more details.}
\end{figure}

For the sake of completeness, we also provide in Figure \ref{fig:other vs T2} the comparison between the THO and well-known estimators of the density derived from bandwidth selectors available in the {\it density} generic function of R. We observe that $\mytilde s_{ucv}$ and $\mytilde s_{SJ}$ perform well (particularly for the $L_1$-loss), whereas the THO outperforms $\mytilde s_{nrd}$ and $\mytilde s_{bcv}$.


\section{Empirical complexity of the exact algorithm}

\begin{figure}[H]
\[ \includegraphics[height=2.5cm,width=0.8\textwidth]{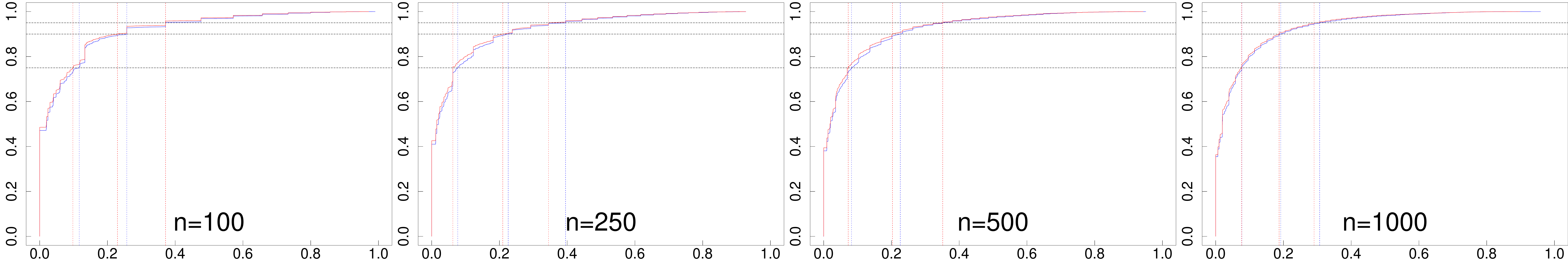} \]
\caption{\label{complexityCDF} \scriptsize From left to right, the CDF for $n=100$, 250, 500 and 1000 of the THO complexity using Algorithm \ref{algo1} in plain line: procedure derived from (\ref{test_Birge}) in blue and from (\ref{test_Baraud}) in red. The horizontal black dotted lines provide the values 0.75, 0.9 and 0.95 and the vertical dotted lines their respective quantiles using the respective colors.}
\end{figure}

To evaluate the complexity of our algorithms let us denote by $N$ the number of tests needed in the computation of the THO for each generated sample of our simulations. As  $N$ is between $M-1$ and $M(M-1)/2$, we define the so-called ``THO complexity'' as the ratio of $N-M+1$ over its maximal value, that is \begin{equation}\frac{2(N - M + 1)}{(M-1)(M-2)}\enspace.\label{eq:complexity}\end{equation} For any run, this ratio belongs to $[0,1]$ by construction. For each fixed $n$,  we get a global sample of size 10800 corresponding to ``18 densities'' times ``6 families'' times ``100 simulations''. Figure \ref{complexityCDF} shows the empirical cumulative distribution function (CDF) of the latter sample with the quantiles 0.75, 0.9 and 0.95, for both tests (\ref{test_Birge}) and (\ref{test_Baraud}). We observe from this figure that in both cases the complexity of our algorithm tends to improve with $n$. Moreover, 75\% of the THO complexities are smaller than 0.1 for $n$ equals 250, 500 and 1000 and 95\% are smaller than 0.4 for all values of $n$. The THO complexity using (\ref{test_Baraud}) is slightly smaller. However the comparison of two estimators in (\ref{test_Baraud}) requires the computation of one integral to compute the difference of squared Hellinger distances involving the middle point. From a practical point-of-view, we indeed observed that using the test (\ref{test_Baraud}) is more CPU time-consuming. Since both strategies have similar THO complexity, we pursue our study again using the procedure derived from (\ref{test_Birge}) only.

\begin{figure}[H]
\[ \includegraphics[height=2.5cm,width=0.4\textwidth]{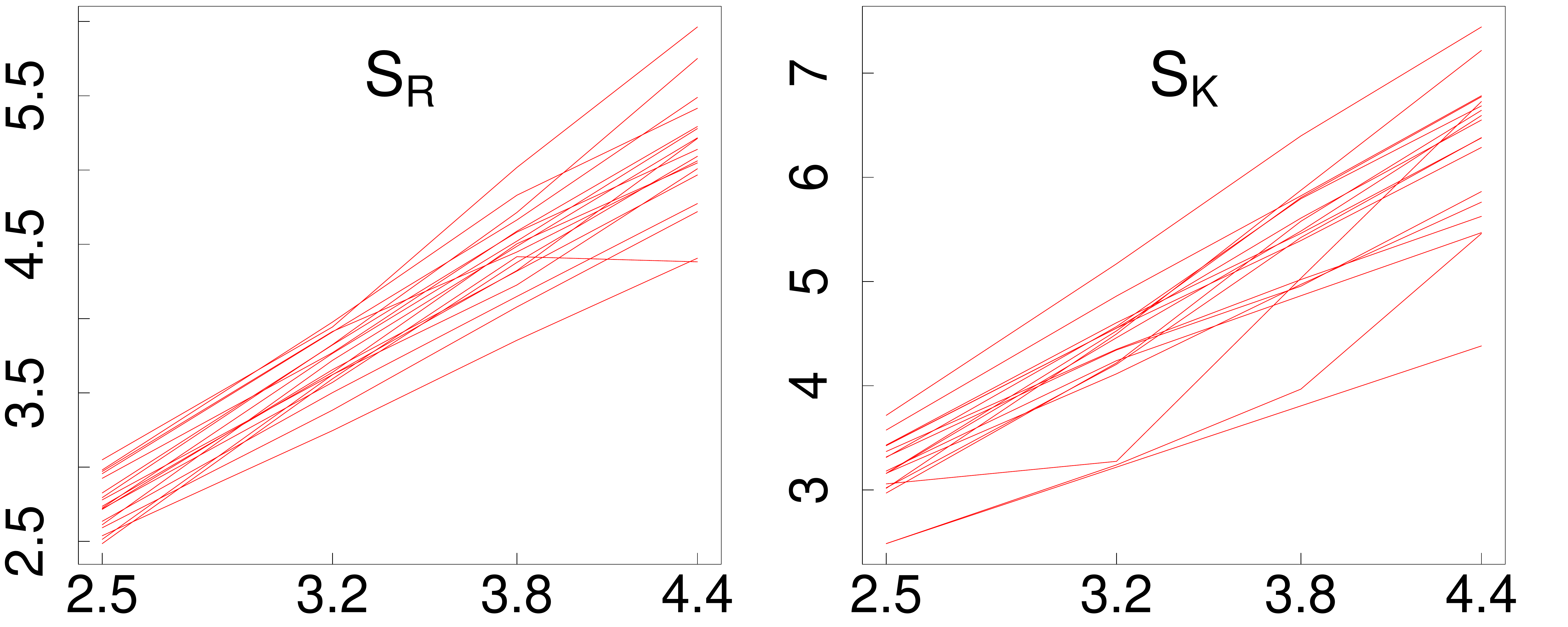} \]
\caption{\label{complexity plot} \scriptsize  Graphs of $\log(N)$ versus $\log(M-1)$ for each density when using the collections $S_R$ (left) and $S_K$ (right).}
\end{figure}

In order to complete this study of the complexity we focused on the two collections $S_R$ and $S_K$ for which the number of estimators depends on $n$ as $M=\lceil n_1/\log(n_1)\rceil$. Having in mind that $N$ is not smaller than $M-1$ and not larger than $M(M-1)/2$, we assumed $N$ to be of order $(M-1)^\beta$ with $\beta$ in $[1,2]$. For each density and each value of $n$, we compute the average of $\log(N)$ over the 100 runs. In Figure \ref{complexity plot} these average values are drawn versus $\log(M-1)$ for the two collections and for each density.  

As Figure \ref{complexity plot} exhibits mostly linear behaviors, we computed the slope in the linear model of $\log(N)$ versus $\log(M-1)$  as an estimator of $\beta$ when $n_1$ varies. 
We observe that this estimator concentrates around respectively 1.2 and 1.4 for the collections $S_R$ and $S_K$ providing a good indicator that our algorithm is typically sub-quadratic. 
The larger value of $\beta$ for the collection $S_K$ may be explained by the fact that, for our set of bandwidths, the kernel estimators may be very similar, inducing a slow decrease of the running intersection $J$ in Algorithm \ref{algo1}. 

\section{Study of the approximate T-Hold-Out}

We provide a comparison of the estimators selected using Algorithms \ref{algo1} and \ref{algo2} respectively, that is the exact T-estimator and its approximate version (denoted here by $\myhat m_T^g$) computed with $\delta_n={c}/\sqrt{|\textbf{X}_v|}$ for different values of $c$. We compare these estimators using the two strategies based on $\textbf{X}_t$ and $\textbf{X}$. 
\begin{figure}[H]
\[ \includegraphics[height=2.5cm,width=\textwidth]{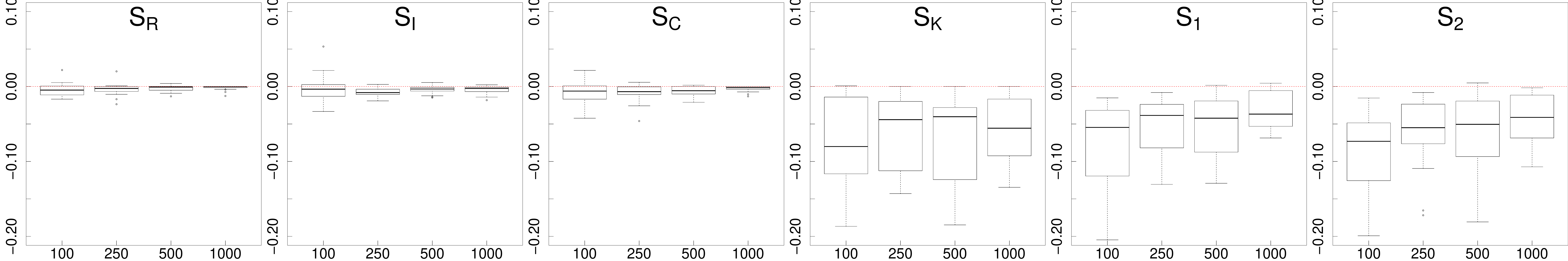} \]
\[ \includegraphics[height=2.5cm,width=\textwidth]{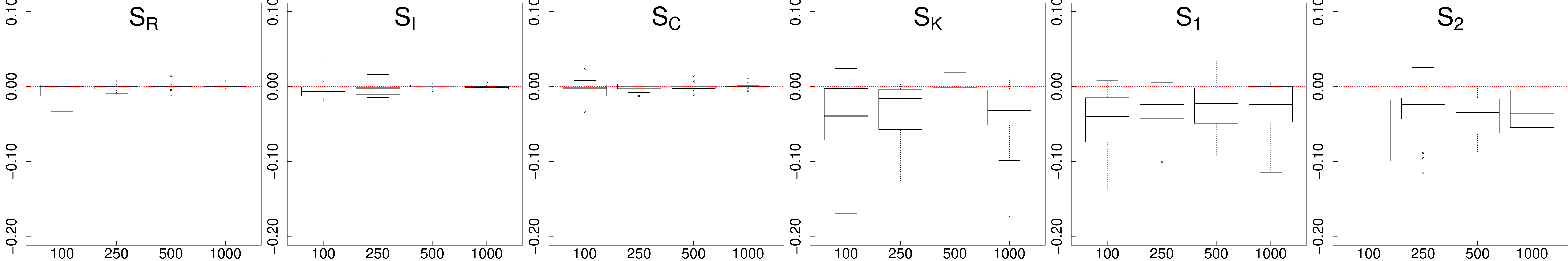} \]
\caption{\label{fig:greedy vs exact} \scriptsize From left to right, normalized $\log_2$-ratio of the empirical risks $\bar{W}_s(\myhat s_{\myhat m_T}[\textbf{X}_t],\myhat s_{\myhat m_T^g}[\textbf{X}_t])$ (upper line) and $\bar{W}_s(\myhat s_{\myhat m_T}[\textbf{X}],\myhat s_{\myhat m_T^g}[\textbf{X}])$ using $c=1$ (bottom line) for the Hellinger loss, using collections $S_R$, $S_I$, $S_C$, $S_K$, $S_1$ and $S_2$. See Figure \ref{fig:p in T1} for more details.}
\end{figure}

\begin{figure}[H]
\[ \includegraphics[height=2.5cm,width=0.8\textwidth]{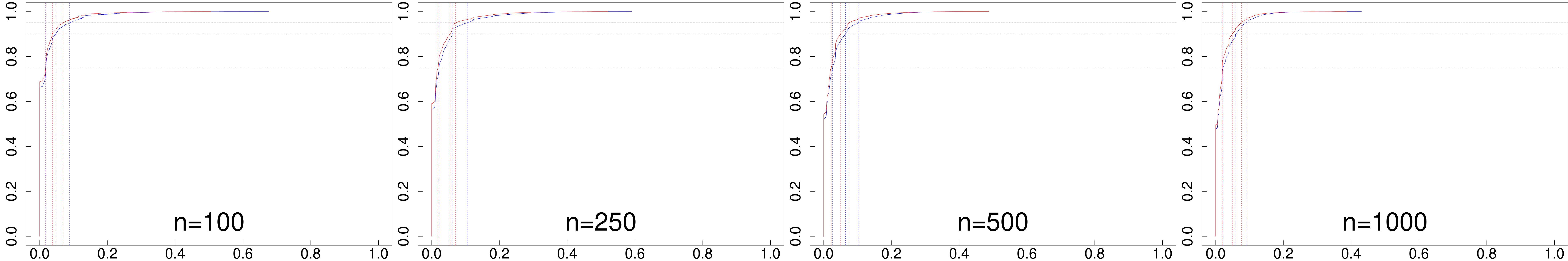} \]
\caption{\label{complexityCDFgreedy} \scriptsize From left to right, the CDF for $n=100$, 250, 500 and 1000 of the THO complexity in plain line using Algorithm \ref{algo2} with $c=1$. See Figure \ref{complexityCDF} for more details.}
\end{figure}

Figure \ref{fig:greedy vs exact} is built using  $\mytilde t_1=\myhat s_{\myhat m_T}[\textbf{X}_t]$ and $\mytilde t_2=\myhat s_{\myhat m_T^g}[\textbf{X}_t]$ with $p=2/3$ on the upper line and using $\mytilde t_1=\myhat s_{\myhat m_T}[\textbf{X}]$ and $\mytilde t_2=\myhat s_{\myhat m_T^g}[\textbf{X}]$ with $p=1/2$ on the bottom line. As expected, the exact THO is better in terms of risk. For histogram families, the degradation of the Hellinger risk is negligible. For families $S_K$, $S_1$ and $S_2$, we observe that the risk increases not more than 20\% in most of the cases ($y$-axis reference value equals to -0.13). The empirical cumulative distribution function (CDF) of the complexity ratio defined in (\ref{eq:complexity}) is shown in Figure \ref{complexityCDFgreedy}, for both tests, for comparison with Figure \ref{complexityCDF}. Clearly the CDFs of the lossy version are more concentrated around 0, showing a significant gain in terms of complexity when using Algorithm \ref{algo2} (quantiles are divided by more than 2.5). 

A further study, using $c=2$ in the approximate algorithm, shows that the risk increases up to 75\% in most of the cases and does not offer a good trade-off between complexity and accuracy.

\section{Conclusion}

We introduce an efficient and exact algorithm, together with an approximate version, for T-estimation in the context of hold-out. We study the performances of this T-hold-out in the density framework using two different robust tests. Calibration study shows that, when building the final estimate only with the training sample, a good choice of the ratio between training and validation sample sizes is $p=2/3$.  However, risks can be improved using the full sample to build the final estimate when using $p=1/2$. Our procedure is competitive compared to classical hold-out derived from Kullback-Leibler or least-squares contrasts. It still behaves well against model selection procedures derived from a calibrated penalized contrast for histogram selection, and against most of the bandwidth selectors for kernel estimators. Empirically, we observe that this algorithm improves clearly the combinatorial complexity. Moreover, it can be speeded up thanks to our proposed lossy version, which offers the expected trade-off between complexity and estimation quality. Finally, the two THO strategies are very similar in terms of Hellinger risk and THO complexity, but we recommend to proceed the THO procedure based on (\ref{test_Birge}) since it is less time-consuming.


\bibliographystyle{plainnat}

\end{document}